\documentclass[aps,pra,preprintnumbers,showpacs,tightenlines]{revtex4}
\usepackage{amssymb}
\usepackage{amsmath}
\usepackage{graphicx}
\usepackage{epsfig}
\usepackage{subfigure}
\usepackage{amsfonts}
\usepackage{CJK}

\begin{document}

\title{Hybrid controlled-SUM gate with one superconducting qutrit and one
cat-state qutrit and application in hybrid entangled state preparation}

\author{Qi-Ping Su$^{1}$}
\author{Yu Zhang$^{2}$}
\author{Liang Bin$^{1}$}
\author{Chui-Ping Yang$^{1,3}$}
\email{yangcp@hznu.edu.cn}

\address{$^1$School of Physics, Hangzhou Normal University, Hangzhou 311121, China }
\address{$^2$School of Physics, Nanjing University, Nanjing 210093, China}
\address{$^3$Quantum Information Research Center, Shangrao Normal University, Shangrao 334001, China}

\date{\today}

\begin{abstract}
Compared with a qubit, a qudit (i.e., $d$-level or $d$-state quantum system) provides a
larger Hilbert space to store and process information. On the other hand,
qudit-based hybrid quantum computing usually requires
performing hybrid quantum gates with qudits different in their nature or in
their encoding format. In this work, we consider the qutrit case, i.e., the case for a qudit with $d$=3.
We propose a simple method to realize a hybrid quantum controlled-SUM gate with one superconducting (SC) qutrit
and a cat-state qutrit. This gate plus single-qutrit gates form a \textit{universal} set of ternary logic gates for quantum computing with qutrits. Our proposal is based on circuit QED and operates essentially by employing a SC ququart (a four-level quantum system) dispersively coupled to a microwave cavity. The gate implementation is quite simple because it only requires a single basic operation. Neither classical pulse nor measurement is needed. The auxiliary higher energy level of the SC ququart is virtually excited during the gate operation, thus decoherence from this level is greatly suppressed. As an application of this gate, we discuss the generation of a hybrid maximally-entangled state of one SC qutrit and one cat-state qutrit. We further analyze the experimental feasibility of creating such hybrid entangled state
in circuit QED. This proposal is quite general and can be extended to accomplish the same task in
a wide range of physical system, such as a four-level natural or artificial
atom coupled to an optical or microwave cavity.
\end{abstract}

\maketitle
\date{\today }

\address{$^1$School of Physics, Hangzhou Normal University, Hangzhou
311121, China }
\address{$^2$School of Physics, Nanjing University, Nanjing
210093, China}
\address{$^3$Quantum Information Research Center, Shangrao
Normal University, Shangrao 334001, China}

\begin{center}
\textbf{I. INTRODUCTION AND MOTIVATION}
\end{center}

Quantum computers are in principle able to solve hard computational problems
much more efficiently than classical computers [1-3]. In the past years,
significant progress has been made in the implementation of quantum
computers, where quantum information is either carried or stored using
qubits, namely two-level or two-state quantum systems. On the other hand,
qudits ($d$-level or $d$-state systems with $d>2$), instead of qubits, are
used to perform quantum computing, which is now a developing field and has
attracted growing attention. Because of its multi-level or multi-state
nature, a qudit has a larger Hilbert space than a qubit. Thus, compared to
their qubit-based counterparts, qudits-based processors can store
exponentially greater information, implement certain algorithms using fewer
entangling gates, and perform more powerful quantum computing [4-8]. Various
physical platforms, such as photonic systems [9,10], ion trap [11],
continuous spin systems [12,13], nitrogen-vacancy centers [14], nuclear
magnetic resonance [6,15], molecular magnets [16], and superconducting
circuits [17-21], have been applied to implement the qudit-based quantum
computing.

Particularly, qutrits (three-level or three-state quantum systems) or qudits
with $d=3$ have been studied both experimentally and theoretically. They
can, in theory, be used for quantum error correction using small code size
[22,23], quantum cryptography [24,25], and efficient communication protocols
[26]. Experimentally, qutrits have been employed for fundamental tests of
quantum mechanics [27] and used as auxiliary systems to accomplish various
quantum computing tasks, such as implementing Toffoli gates [28] and
multiqubit controlled-phase gates [29]. Moreover, quantum teleportations
based on qutrits have been performed in photonic platforms [30,31]. As
relevant to this work, experiments have demonstrated coherent population
transfer in a superconducting qutrit [32], creation of GHZ entangled states
of superconducting qutrits [20,21], and realization of single-qutrit quantum
gates and a two-qutrit controlled-SUM (CSUM) gate with superconducting
qutrits [18-20].

On the other hand, hybrid gates have attracted tremendous attention
recently, because of their importance in connecting quantum information
processors with different encoding information-processing units, as well as
their significant application in transferring quantum states between a
quantum processor and a quantum memory. A number of works on implementing
hybrid gates with various qubits, qutrits or qudits have been presented
(e.g., [33-43]). However, after a deep search of literature, we note that
how to implement hybrid quantum gates with SC qutrits and cat-state qutrits
(i.e., qutrits encoded via cat states) has not been studied yet.

In the following, we will propose a simple method to realize a hybrid
quantum controlled-SUM (CSUM) gate with a three-level SC qutrit and a
cat-state qutrit, based on circuit QED. The circuit QED, composed of
microwave cavities and SC qubits or qudits, has appeared as one of the most
promising candidates for quantum computing [44-51]. SC qubits or qudits,
such as charge, flux, phase, transmon, and Xmon, can be fabricated using
modern integrated circuit technology, their properties can be characterized
and adjusted in situ, they have relatively long decoherence times [52-55],
and thus they are good information processing units in quantum computing. In
addition, the cat-state encoding, consisting of superpositions of coherent
states, is protected against photon loss and dephasing errors [56,57], and
quantum computing based on cat-state encoding has attracted much attention
recently [58-60].

This proposal operates essentially by employing a SC ququart dispersively
coupled to a microwave cavity. Here, \textquotedblleft ququart" refers to a
four-level quantum system, with three levels representing a qutrit and the
fourth level acting as an auxiliary level for the state manipulation. The
gate realization is quite simple because it only requires a single basic
operation. As an application of this gate, we discuss the generation of a
hybrid maximally-entangled state of a SC qutrit and a cat-state qutrit. We
further analyze the circuit-QED experimental feasibility of creating such
hybrid entangled state. This proposal is quite general and can be extended
to accomplish the same task in a wide range of physical system, such as a
four-level natural or artificial atom coupled to an optical or microwave
cavity.

Other motivations of this work are as follows. First, hybrid gates of SC
qutrits and cat-state qutrits are of significance in realizing a large-scale
quantum computing executed in a compounded information processor, which is
composed of a SC-qutrit based quantum processor and a cat-state-qutrit based
quantum processor. Second, they are also important in the transmission of
quantum states between a SC-qutrit based quantum processor and a
cat-state-qutrit based quantum memory [61]. Cat-state qutrits could be good
memory units for storing high-dimensional quantum states because experiments
have demonstrated that the lifetime of storing quantum information via
cat-state encoding can be greatly enhanced through quantum error correction
[62]. Last, a two-qutrit CSUM gate plus single-qutrit gates form a \textit{%
universal} set of ternary logic gates for quantum computing with qutrits
[63-65], and thus implementing a hybrid two-qutrit CSUM gate becomes
necessary in hybrid quantum computing with qutrits.

This paper is organized as follows. In Sec. II, we briefly introduce a
hybrid two-qutrit CSUM gate. In Sec. III, we explicitly show how to realize
this hybrid gate with a SC qutrit and a cat-state qutrit. In Sec. IV, we
show how to generate a hybrid maximally-entangled state of a SC qutrit and a
cat-state qutrit by applying this gate. In Sec. V, we give a discussion on
the experimental feasibility of creating this hybrid entangled state, by
employing a SC flux ququart coupled to a three-dimensional (3D) microwave
cavity. A concluding summary is given in Sec. VI.

\begin{center}
\textbf{II. HYBRID TWO-QUTRIT CSUM GATE}
\end{center}

A qutrit has three logic states, which are denoted as $\left\vert
0\right\rangle ,$ $\left\vert 1\right\rangle ,$ and $\left\vert
2\right\rangle $, respectively. For two qutrits, there are a total number of
nine computational basis states, i.e., $\left\vert 00\right\rangle
,\left\vert 01\right\rangle ,\left\vert 02\right\rangle ,\left\vert
10\right\rangle ,\left\vert 11\right\rangle ,\left\vert 12\right\rangle
,\left\vert 20\right\rangle ,\left\vert 21\right\rangle ,$and $\left\vert
22\right\rangle $. A two-qutrit CSUM gate is described by
\begin{eqnarray}
\left\vert 00\right\rangle &\rightarrow &\left\vert 00\right\rangle ,\text{ }%
\left\vert 01\right\rangle \rightarrow \left\vert 01\right\rangle ,\text{ }%
\left\vert 02\right\rangle \rightarrow \left\vert 02\right\rangle ,  \notag
\\
\left\vert 10\right\rangle &\rightarrow &\left\vert 11\right\rangle ,\text{ }%
\left\vert 11\right\rangle \rightarrow \left\vert 12\right\rangle ,\text{ }%
\left\vert 12\right\rangle \rightarrow \left\vert 10\right\rangle ,  \notag
\\
\left\vert 20\right\rangle &\rightarrow &\left\vert 22\right\rangle ,\text{ }%
\left\vert 21\right\rangle \rightarrow \left\vert 20\right\rangle ,\text{ }%
\left\vert 22\right\rangle \rightarrow \left\vert 21\right\rangle ,
\end{eqnarray}%
which implies that: (i) if the control qutrit (the first qutrit) is in the
state $\left\vert 1\right\rangle ,$ the state $\left\vert k\right\rangle $
of the target qutrit (the second qutrit)\ is shifted to the state $%
\left\vert k\oplus 1\right\rangle ;$ (ii) if the control qutrit (the first
qutrit) is in the state $\left\vert 2\right\rangle ,$ the state $\left\vert
k\right\rangle $ of the target qutrit is shifted to the state $\left\vert
k\oplus 2\right\rangle ;$ however (iii) when the control qutrit is in the
state $\left\vert 0\right\rangle ,$ the state $\left\vert k\right\rangle $
of the target qutrit remains unchanged. Here, $k\in \{0,1,2\},$ $\left\vert
k\oplus 1\right\rangle $ means $k+1$ mod 3, while $\left\vert k\oplus
2\right\rangle $ represents $k+2$ mod 3.

The hybrid two-qutrit CSUM gate considered in this work is described by Eq.
(1). The control qutrit is a SC qutrit, whose three logic states are
represented by the three lowest levels $\left\vert 0\right\rangle ,$ $%
\left\vert 1\right\rangle ,$ and $\left\vert 2\right\rangle $ of a SC ququart%
\textrm{\ }(Fig. 1a), while the target qutrit is a cat-state qutrit, for
which the three logic states $\left\vert 0\right\rangle ,$\textrm{\ }$%
\left\vert 1\right\rangle ,$\textrm{\ and }$\left\vert 2\right\rangle $ are
encoded via three quasi-orthogonal cat states of a cavity, which are given
below
\begin{eqnarray}
\left\vert 0\right\rangle &=&\mathcal{M}_{0}\ \left( |\alpha \rangle
+|-\alpha \rangle \right) ,  \notag \\
\left\vert 1\right\rangle &=&\mathcal{M}_{1}\ \left( |\alpha e^{i\pi
/3}\rangle +|-\alpha e^{i\pi /3}\rangle \right) ,  \notag \\
\left\vert 2\right\rangle &=&\mathcal{M}_{2}\ \left( |\alpha e^{i2\pi
/3}\rangle +|-\alpha e^{i2\pi /3}\rangle \right) .
\end{eqnarray}%
Here, $\mathcal{M}_{0},$ $\mathcal{M}_{1},$ and $\mathcal{M}_{2}$\ are
normalization coefficients, with $\mathcal{M}_{0}=$ $\mathcal{M}_{1}=%
\mathcal{M}_{2}=1/\sqrt{2\left( 1+e^{-2\left\vert \alpha \right\vert
^{2}}\right) }.$ For $\alpha \geq 3.05,$ one can verify $\left\vert
\left\langle k\right\vert \left. l\right\rangle \right\vert ^{2}<10^{-4}$
for $k\neq l$ ($k,l\in \left\{ 0,1,2\right\} $). Thus, when $\alpha $ is
large enough, any two of the three logic states of the cat-state qutrit can
be made to be quasi-orthogonal to each other.

In the next section, we will show how to realize this hybrid CSUM gate. To
avoid the confusion, please keep in mind that for each one of the states $%
\left\vert 00\right\rangle ,\left\vert 01\right\rangle ,\left\vert
02\right\rangle ,\left\vert 10\right\rangle ,\left\vert 11\right\rangle
,\left\vert 12\right\rangle ,\left\vert 20\right\rangle ,\left\vert
21\right\rangle ,$ and $\left\vert 22\right\rangle $ listed below, the left
0, 1, and 2 correspond to the SC qutrit, while the right 0, 1, and 2
correspond to the cat-state qutrit.

\begin{center}
\textbf{III. IMPLEMENTING A HYBRID TWO-QUTRIT CSUM GATE}
\end{center}

Let us now consider a setup consisting of a microwave cavity and a SC
ququart (Figs.~1b and 1c). The SC ququart has three lowest levels $|0\rangle
$, $|1\rangle ,|2\rangle $\ and an auxiliary higher-energy level $\left\vert
b\right\rangle $ (Fig. 1a). The SC ququart is initially decoupled from the
cavity. The procedure for implementing the gate is listed below:

\begin{figure}[tbp]
\begin{center}
\includegraphics[bb=43 176 900 411, width=12.5 cm, clip]{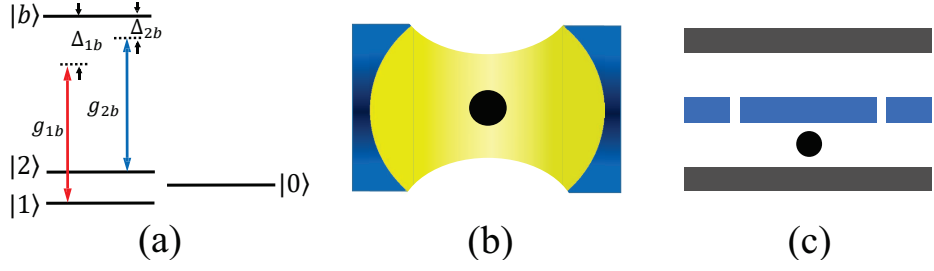} \vspace*{%
-0.08in}
\end{center}
\caption{(color online) (a) The cavity is dispersively coupled to the $%
|1\rangle \leftrightarrow |b\rangle $\ transition of the SC ququart with
coupling constant $g_{1b}$\ and detuning $\Delta _{1b}$, while dispersively
coupled to the $|2\rangle \leftrightarrow |b\rangle $\ transition of the SC
ququart with coupling constant $g_{2b}$\ and detuning $\Delta _{2b}$. (b)
Setup for one SC ququart embedded in a 3D microwave cavity. (c) Setup for a
SC ququart coupled to a 1D microwave cavity or resonator. In (b) and (c),
the dark dot represents the SC ququart.}
\label{fig:1}
\end{figure}

Adjust the cavity frequency such that the cavity is dispersively coupled to
the $|1\rangle \leftrightarrow |b\rangle $\ transition of the SC ququart
with coupling constant $g_{1b}$\ and detuning $\Delta _{1b}$, the cavity is
dispersively coupled to the $|2\rangle \leftrightarrow |b\rangle $\
transition of the SC ququart with coupling constant $g_{2b}$\ and detuning $%
\Delta _{2b}$, but highly detuned (decoupled) from the transition between
any other two levels of the SC ququart (Fig.~1a). Under these
considerations, the Hamiltonian in the interaction picture is thus given by
(in units of $\hbar =1$)
\begin{equation}
H_{\mathrm{I}}=g_{1b}(e^{i\Delta _{1b}t}\hat{a}|b\rangle \langle 1|+\text{%
h.c.})+g_{2b}(e^{i\Delta _{2b}t}\hat{a}|b\rangle \langle 2|+\text{h.c.}),
\end{equation}%
where $\hat{a}$\ is the photon annihilation operator of the cavity, $\Delta
_{1b}=\omega _{1b}-\omega _{c},$\ and $\Delta _{2b}=\omega _{2b}-\omega _{c}$%
. Here, $\omega _{1b}$ ($\omega _{2b}$) is the $|1\rangle \leftrightarrow
|b\rangle $\ $\left( |2\rangle \leftrightarrow |b\rangle \right) $
transition frequency of the ququart, while $\omega _{c}$ is the cavity
frequency.

In the large-detuning regime of $\left\vert \Delta _{1b}\right\vert \gg
g_{1b}$ and $\left\vert \Delta _{2b}\right\vert \gg g_{2b}$, the Hamiltonian
(3) becomes [66-68]
\begin{equation}
H_{\mathrm{e}}=-\lambda _{1}(\hat{a}^{+}\hat{a}|1\rangle \langle 1|-\hat{a}%
\hat{a}^{+}|b\rangle \left\langle b\right\vert )-\lambda _{2}(\hat{a}^{+}%
\hat{a}|2\rangle \langle 2|-\hat{a}\hat{a}^{+}|b\rangle \left\langle
b\right\vert ),
\end{equation}%
where $\lambda _{1}=g_{1b}^{2}/\Delta _{1b},$ $\lambda
_{2}=g_{2b}^{2}/\Delta _{2b},$ the terms in the first (second) bracket
describe the photon-number dependent stark shifts of the energy levels $%
|1\rangle $ ($\left\vert 2\right\rangle $) and $|b\rangle ,$ which are
induced by the cavity.

Note that the Hamiltonian (4) does not induce the transition between any two
of the four levels $|0\rangle $, $|1\rangle ,$ $|2\rangle ,$ and $|b\rangle $%
. Hence, as long as the auxiliary level $|b\rangle $\ is initially
unpopulated, this level will remain unoccupied. Therefore, the Hamiltonian
(4) reduces to
\begin{equation}
\widetilde{H}_{\mathrm{e}}=-\lambda _{1}\hat{a}^{+}\hat{a}|1\rangle \langle
1|-\lambda _{2}\hat{a}^{+}\hat{a}|2\rangle \langle 2|.
\end{equation}%
For this Hamiltonian, the unitary operator $U=e^{-i\widetilde{H}_{\mathrm{e}%
}t}$ results in the following state transformation%
\begin{eqnarray}
\left\vert 10\right\rangle &\rightarrow &\left\vert 1\right\rangle \mathcal{M%
}_{0}\left( |\alpha e^{i\lambda _{1}t}\rangle +|-\alpha e^{i\lambda
_{1}t}\rangle \right) ,\text{ \ }  \notag \\
\left\vert 11\right\rangle &\rightarrow &\left\vert 1\right\rangle \mathcal{M%
}_{1}\left( |\alpha e^{i\pi /3}e^{i\lambda _{1}t}\rangle +|-\alpha e^{i\pi
/3}e^{i\lambda _{1}t}\rangle \right) ,\text{\ }  \notag \\
\left\vert 12\right\rangle &\rightarrow &\left\vert 1\right\rangle \mathcal{M%
}_{2}\left( |\alpha e^{i2\pi /3}e^{i\lambda _{1}t}\rangle +|-\alpha e^{i2\pi
/3}e^{i\lambda _{1}t}\rangle \right) ,\text{\ \ \ }  \notag \\
\left\vert 20\right\rangle &\rightarrow &\left\vert 2\right\rangle \mathcal{M%
}_{0}\left( |\alpha e^{i\lambda _{2}t}\rangle +|-\alpha e^{i\lambda
_{2}t}\rangle \right) ,\text{ \ }  \notag \\
\left\vert 21\right\rangle &\rightarrow &\left\vert 2\right\rangle \mathcal{M%
}_{1}\left( |\alpha e^{i\pi /3}e^{i\lambda _{2}t}\rangle +|-\alpha e^{i\pi
/3}e^{i\lambda _{2}t}\rangle \right) ,\text{ }  \notag \\
\left\vert 22\right\rangle &\rightarrow &\left\vert 2\right\rangle \mathcal{M%
}_{2}\left( |\alpha e^{i2\pi /3}e^{i\lambda _{2}t}\rangle +|-\alpha e^{i2\pi
/3}e^{i\lambda _{2}t}\rangle \right) ,
\end{eqnarray}%
where on the left side the three logic states $\left\vert 0\right\rangle ,$ $%
\left\vert 1\right\rangle $, and $\left\vert 2\right\rangle $ of the
cat-state qutrit are defined in Eq. (2) above..

If we set
\begin{equation}
\lambda _{1}t=\pi /3,\text{ }\lambda _{2}t=2\pi /3,\text{ }
\end{equation}%
(i.e., $\lambda _{2}=2\lambda _{1}$), then the state transformation (6)
becomes
\begin{eqnarray}
\left\vert 10\right\rangle &\rightarrow &\left\vert 1\right\rangle \mathcal{M%
}_{0}\left( |\alpha e^{i\pi /3}\rangle +|-\alpha e^{i\pi /3}\rangle \right) ,%
\text{ }  \notag \\
\left\vert 11\right\rangle &\rightarrow &\left\vert 1\right\rangle \mathcal{M%
}_{1}\left( |\alpha e^{i2\pi /3}\rangle +|-\alpha e^{i2\pi /3}\rangle
\right) ,\text{ }  \notag \\
\left\vert 12\right\rangle &\rightarrow &\left\vert 1\right\rangle \mathcal{M%
}_{2}\left( |\alpha \rangle +|-\alpha \rangle \right) ,  \notag \\
\left\vert 20\right\rangle &\rightarrow &\left\vert 2\right\rangle \mathcal{M%
}_{0}\left( |\alpha e^{i2\pi /3}\rangle +|-\alpha e^{i2\pi /3}\rangle
\right) ,\text{ }  \notag \\
\left\vert 21\right\rangle &\rightarrow &\left\vert 2\right\rangle \mathcal{M%
}_{1}\left( |\alpha \rangle +|-\alpha \rangle \right) ,\text{ \ \ \ \ \ \ \
\ }  \notag \\
\left\vert 22\right\rangle &\rightarrow &\left\vert 2\right\rangle \mathcal{M%
}_{2}\left( |\alpha e^{i\pi /3}\rangle +|-\alpha e^{i\pi /3}\rangle \right) .
\end{eqnarray}

According to the three logic states of the cat-state qutrit defined in
Eq.~(2) and because of $\mathcal{M}_{0}=$ $\mathcal{M}_{1}=\mathcal{M}_{2}$,
the state transformation (8) can be written as
\begin{eqnarray}
\left\vert 10\right\rangle &\rightarrow &\left\vert 11\right\rangle ,\text{ }%
\left\vert 11\right\rangle \rightarrow \left\vert 12\right\rangle ,\text{ }%
\left\vert 12\right\rangle \rightarrow \left\vert 10\right\rangle ,\text{\ }
\notag \\
\left\vert 20\right\rangle &\rightarrow &\left\vert 22\right\rangle ,\text{\
}\left\vert 21\right\rangle \rightarrow \left\vert 20\right\rangle ,\text{ }%
\left\vert 22\right\rangle \rightarrow \left\vert 21\right\rangle .
\end{eqnarray}

On the other hand, it is noted that the Hamiltonian (5) does not involve the
level $\left\vert 0\right\rangle $ of the SC ququart, hence this Hamiltonian
(5) acting on the three states $\left\vert 00\right\rangle ,\left\vert
01\right\rangle ,$and $\left\vert 02\right\rangle $ results in zero, i.e., $%
\widetilde{H}_{\mathrm{e}}\left\vert 00\right\rangle =\widetilde{H}_{\mathrm{%
e}}\left\vert 01\right\rangle =\widetilde{H}_{\mathrm{e}}\left\vert
02\right\rangle \equiv 0.$ As a result, the other three states $\left\vert
00\right\rangle ,\left\vert 01\right\rangle ,\left\vert 02\right\rangle $ of
the two qutrits remain unchanged under the unitary operator $U=e^{-i%
\widetilde{H}_{\mathrm{e}}t},$ i.e.,%
\begin{equation}
\left\vert 00\right\rangle \rightarrow \left\vert 00\right\rangle ,\text{ }%
\left\vert 01\right\rangle \rightarrow \left\vert 01\right\rangle ,\text{ }%
\left\vert 02\right\rangle \rightarrow \left\vert 02\right\rangle .\text{\ }
\end{equation}%
Therefore, it can be concluded from Eqs.~(9) and (10) that the hybrid
two-qutrit CSUM gate described by Eq.~(1) is implemented with a SC qutrit
(the control qutrit) and a cat-state qutrit (the target qutrit) after the
above operation. It should be noted that after the operation, the cavity
frequency needs to be adjusted back such that the cavity is decoupled from
the SC ququart.

In the above, we have set $\lambda _{2}=2\lambda _{1}$, which turns out into
\begin{equation}
g_{2b}^{2}/\Delta _{2b}=2g_{1b}^{2}/\Delta _{1b}.
\end{equation}%
Note that this condition (11) can be readily satisfied by carefully
selecting $g_{1b},g_{2b},\Delta _{1b}$ or $\Delta _{2b}.$

During the gate operation described above, the coupling or decoupling of the
SC ququart with the cavity is realized by adjusting the cavity frequency.
For a superconducting microwave cavity, the cavity frequency can be rapidly
(within a few nanoseconds) tuned in experiments [69-71]. Alternatively, the
coupling or decoupling of the SC ququart with the cavity can be obtained by
adjusting the level spacings of the SC ququart. For superconducting qubits
or qudits, their level spacings can be rapidly (within 1--3 ns) adjusted by
varying external control parameters [72,73].

As shown above, the gate realization requires only a single basic operation
described by $U$. During the gate operation, the auxiliary higher energy
level $\left\vert b\right\rangle $ of the ququart is virtually excited and
thus decoherence from this level is greatly suppressed. Moreover, neither
applying a classical pulse to the qutrit nor making measurement on the state
of the SC ququart or the cavity is needed.

\begin{center}
\textbf{IV. PREPARING A HYBRID MAXIMALLY-ENTANGLED STATE OF TWO QUTRITS}
\end{center}

Hybrid entangled states play a crucial role in quantum information
processing and quantum technology. For instance, hybrid entangled states can
be used as quantum channels and intermediate resources for quantum
technology, including quantum information transmission, quantum state
operation, and storage between different encodings and formats [74-76]. They
can also act as practical interfaces to connect quantum processors with
information processing units of different encoding. Generally speaking,
hybrid entangled states involve subsystems different in their nature (e.g.,
photons and matters) or in the degree of freedom (e.g., discrete-variable
degree and continuous variable degree).

Assume that the SC ququart is in a superposition state $\left( \left\vert
0\right\rangle +\left\vert 1\right\rangle +\left\vert 2\right\rangle \right)
/\sqrt{3},$ which can be easily prepared by applying resonant classical
pulses to the SC ququart. Namely, apply a classical pulse (with an initial
phase $\phi =-\pi /2,$ a duration $\tau =\Omega _{1}^{-1}\arccos 1/\sqrt{3},$
resonant to the $\left\vert 0\right\rangle \leftrightarrow \left\vert
1\right\rangle $ transition of the ququart in the state $\left\vert
0\right\rangle $) to achieve the state transformation $\left\vert
0\right\rangle \rightarrow 1/\sqrt{3}\left\vert 0\right\rangle +\sqrt{2/3}%
\left\vert 1\right\rangle $ [77]$;$ then apply a second classical pulse
(with an initial phase $\phi =-\pi /2,$ a duration $\tau =\Omega
_{2}^{-1}\pi /4,$ resonant to the $\left\vert 1\right\rangle \leftrightarrow
\left\vert 2\right\rangle $ transition of the ququart) to achieve the state
transformation $\left\vert 1\right\rangle \rightarrow \left( \left\vert
1\right\rangle +\left\vert 2\right\rangle \right) /\sqrt{2}.$ Here, $\Omega
_{1}$ ($\Omega _{2}$) is the Rabi frequency of the first (second) pulse. It
is easy to check that the superposition state $\left( \left\vert
0\right\rangle +\left\vert 1\right\rangle +\left\vert 2\right\rangle \right)
/\sqrt{3}$ of the ququart is prepared by combination of these two state
transformations.

Suppose that the cavity is in the cat state $\left\vert 0\right\rangle =%
\mathcal{M}_{0}\left( \left\vert \alpha \right\rangle +\left\vert -\alpha
\right\rangle \right) $. Note that this cat state has been experimentally
created in circuit QED [78-82]. The initial state of the system is thus
given by
\begin{equation}
\left\vert \psi \left( 0\right) \right\rangle =\frac{1}{\sqrt{3}}\left(
\left\vert 0\right\rangle +\left\vert 1\right\rangle +\left\vert
2\right\rangle \right) \left\vert 0\right\rangle .
\end{equation}%
Now we perform the operation described in the preceding section to achieve a
hybrid\ two-qutrit CSUM gate (1). From Eq.~(1), one can see that after this
gate operation, the state (12) becomes
\begin{equation}
\frac{1}{\sqrt{3}}\left( \left\vert 00\right\rangle +\left\vert
11\right\rangle +\left\vert 22\right\rangle \right) ,
\end{equation}%
which is the hybrid maximally-entangled state of a SC qutrit and a cat-state
qutrit.

As can be seen from the above description, applying the hybrid two-qutrit
CSUM gate (1) can directly create the hybrid maximally-entangled state (13)\
of a SC qutrit and a cat-state qutrit. Given that the initial state (12) of
the system is ready, the operation time for preparing the hybrid entangled
state (13) is equal to that for implementing the gate (1), i.e., $t=\pi
/\left( 3\lambda _{1}\right) $. Since the hybrid entangled state here is
created based on the gate (1), the Hamiltonian used for generating the
hybrid entangled state (13) is the same as that used for realizing the gate
(1).

\begin{center}
\textbf{IV. EXPERIMENTAL FEASIBILITY}
\end{center}

As an example, let us now give a brief discussion on the possibility of
experimentally creating the hybrid entangled state (13), by considering a
setup of a SC flux ququart coupled to a 3D microwave cavity (Fig.~1b). The
SC flux ququart has four levels as depicted in Fig. 2a, where the \textit{%
ground} level is labelled by $\left\vert 1\right\rangle $. Note that the
transition between the level $\left\vert 0\right\rangle $ and any one of the
other three levels $\left\{ \left\vert 1\right\rangle ,\left\vert
2\right\rangle ,\left\vert b\right\rangle \right\} $ can be made weak by
increasing the barrier between the two potential wells.

As shown in the previous section, the hybrid entangled state (13) was
created by applying the hybrid CSUM gate (1). In this sense, as long as the
initial state (12) can be well prepared, the operational fidelity for the
preparation of the hybrid entangled state (13) depends mainly on the
performance of the hybrid CSUM gate (1) on a SC qutrit and a cat-state
qutrit.

\begin{center}
\textbf{A. Full Hamiltonian}
\end{center}

The hybrid CSUM gate (1) was implemented based on the effective Hamiltonian
(5), which was derived starting from the original Hamiltonian (3). Note that
the Hamiltonian (3) only contains the coupling of the cavity with the $%
|1\rangle \leftrightarrow |b\rangle $\ transition and the coupling of the
cavity with the $|2\rangle \leftrightarrow |b\rangle $ transition of the SC
flux ququart. In reality, there exist the unwanted couplings of the cavity
with other intra-level transitions of the SC flux ququart.

\begin{figure}[tbp]
\begin{center}
\includegraphics[bb=42 87 900 481, width=12.5 cm, clip]{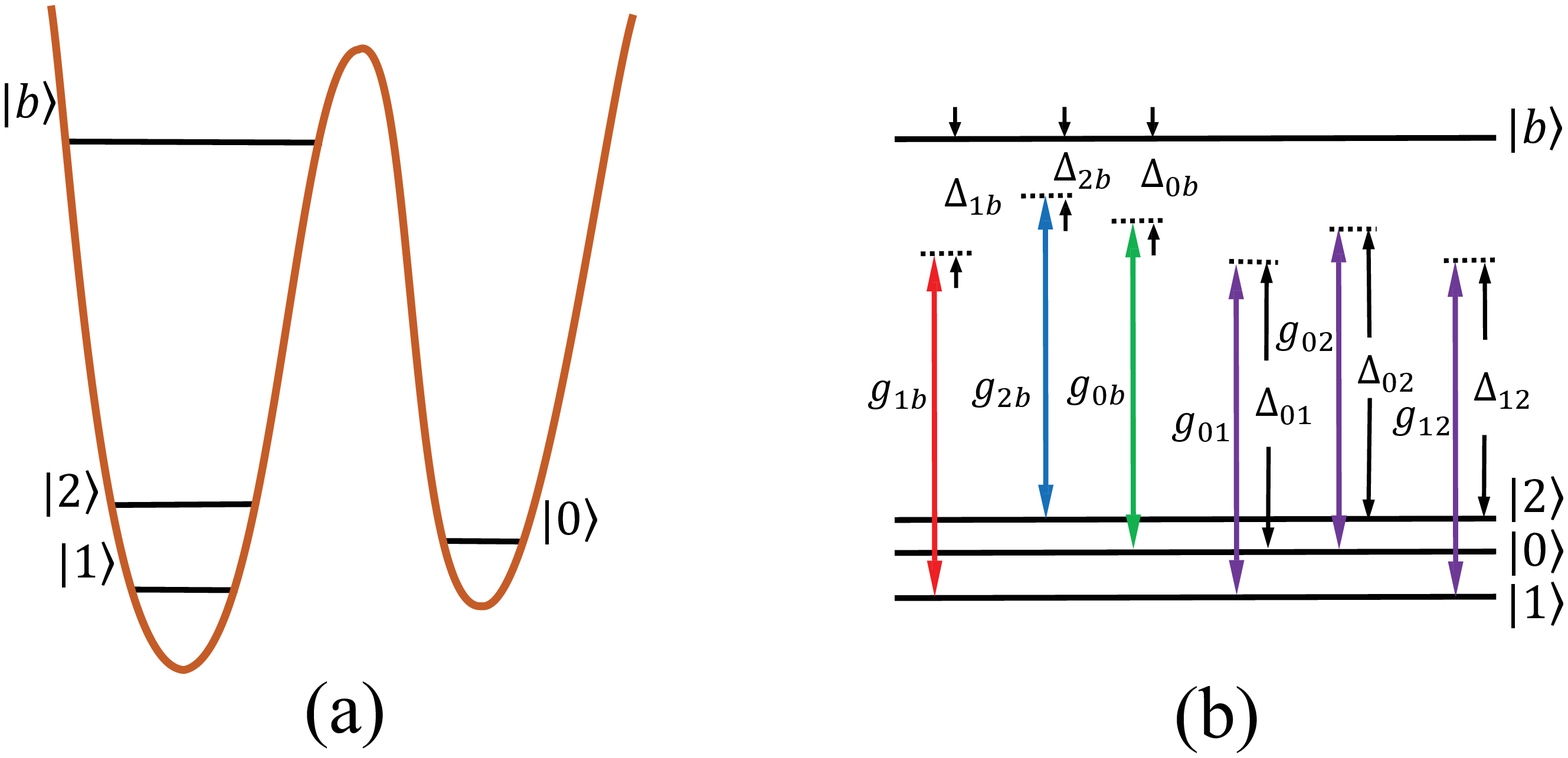} \vspace*{%
-0.08in}
\end{center}
\caption{(color online) (a) Illustration of the four levels of a SC flux
ququart. Note that the \textit{ground} level is labelled by $|1\rangle$. The
transition between the level $\left\vert 0\right\rangle $ and any one of the
other three levels $\left\{ \left\vert 1\right\rangle ,\left\vert
2\right\rangle ,\left\vert b\right\rangle \right\} $ is weak due to the
barrier between the two potential wells. (b) Illustration of the cavity
being dispersively coupled to the $|k\rangle \leftrightarrow |b\rangle $\
transition of the SC ququart with coupling constant $g_{kb}$\ and detuning $%
\Delta _{kb}$ ($k=1,2$), the unwanted coupling between the cavity and the $%
|0\rangle \leftrightarrow |b\rangle $\ transition of the SC ququart with
coupling constant $g_{0b}$\ and detuning $\Delta _{0b}$, as well as the
unwanted coupling between the cavity and the $|m\rangle \leftrightarrow
|n\rangle $\ transition of the SC ququart with coupling constant $g_{mn}$\
and detuning $\Delta _{mn}$ ($mn=01,02,12$). }
\label{fig:2}
\end{figure}

By taking the unwanted couplings into account, the Hamiltonian (3) is
modified as (without RWA)
\begin{eqnarray}
H_{\mathrm{I}}^{\prime } &=&g_{1b}(e^{i\Delta _{1b}t}\hat{a}|b\rangle
\langle 1|+\text{h.c.})+g_{1b}\left[ e^{i\left( \omega _{c}+\omega
_{1b}\right) t}\hat{a}^{+}|b\rangle \langle 1|+\text{h.c.}\right]   \notag \\
&&+g_{2b}(e^{i\Delta _{2b}t}\hat{a}|b\rangle \langle 2|+\text{h.c.})+g_{2b}%
\left[ e^{i\left( \omega _{c}+\omega _{2b}\right) t}\hat{a}^{+}|b\rangle
\langle 2|+\text{h.c.}\right]   \notag \\
&&+g_{0b}(e^{i\Delta _{0b}t}\hat{a}|b\rangle \langle 0|+\text{h.c.})+g_{0b}%
\left[ e^{i\left( \omega _{c}+\omega _{0b}\right) t}\hat{a}^{+}|b\rangle
\langle 0|+\text{h.c.}\right]   \notag \\
&&+g_{01}(e^{i\Delta _{01}t}\hat{a}|0\rangle \langle 1|+\text{h.c.})+g_{01}%
\left[ e^{i\left( \omega _{c}+\omega _{01}\right) t}\hat{a}^{+}|0\rangle
\langle 1|+\text{h.c.}\right]   \notag \\
&&+g_{02}(e^{i\Delta _{02}t}\hat{a}|2\rangle \langle 0|+\text{h.c.})+g_{02}%
\left[ e^{i\left( \omega _{c}+\omega _{02}\right) t}\hat{a}^{+}|2\rangle
\langle 0|+\text{h.c.}\right]   \notag \\
&&+g_{12}(e^{i\Delta _{12}t}\hat{a}|2\rangle \langle 1|+\text{h.c.})+g_{12}%
\left[ e^{i\left( \omega _{c}+\omega _{12}\right) t}\hat{a}^{+}|2\rangle
\langle 1|+\text{h.c.}\right] ,
\end{eqnarray}%
where the terms in the first line correspond to the coupling of the cavity
with the $|1\rangle \leftrightarrow |b\rangle $ transition of the SC
ququart, the terms in the second line correspond to the coupling of the
cavity with the $|2\rangle \leftrightarrow |b\rangle $ transition of the SC
ququart, the terms in the third line correspond to the unwanted coupling
between the cavity and the $|0\rangle \leftrightarrow |b\rangle $ transition
of the SC ququart with coupling constant $g_{0b}$, the terms in the fourth
line correspond to the unwanted coupling between the cavity and the $%
|0\rangle \leftrightarrow |1\rangle $ transition of the SC ququart with
coupling constant $g_{01}$, the terms in the fifth line correspond to the
unwanted coupling between the cavity and the $|0\rangle \leftrightarrow
|2\rangle $ transition of the SC ququart with coupling constant $g_{02}$,
and the terms in the last line correspond to the unwanted coupling between
the cavity and the $|1\rangle \leftrightarrow |2\rangle $ transition of the
SC ququart with coupling constant $g_{12}$ (Fig.~2b). In Eq. (14), $\Delta
_{0b},$ $\Delta _{01},$ $\Delta _{02},$ and $\Delta _{12}$ are detunings,
which are given by $\Delta _{0b}=\omega _{0b}-\omega _{c},$ $\Delta
_{01}=\omega _{01}-\omega _{c},$ $\Delta _{02}=\omega _{02}-\omega _{c},$
and $\Delta _{12}=\omega _{12}-\omega _{c}$ (Fig.~2b). Here, $\omega
_{0b},\omega _{01},\omega _{02},$ and $\omega _{12}$ are the $|0\rangle
\leftrightarrow |b\rangle ,$\ $|0\rangle \leftrightarrow |1\rangle
,|0\rangle \leftrightarrow |2\rangle ,$ and $|1\rangle \leftrightarrow
|2\rangle $ transition frequencies of the ququart, respectively.

\begin{center}
\textbf{B. Numerical results}
\end{center}

The dynamics of the lossy system, with ququart relaxation, dephasing and
cavity decay being included, is determined by%
\begin{eqnarray}
\frac{d\rho }{dt} &=&-i\left[ H_{\mathrm{I}}^{\prime },\rho \right] +\kappa
\mathcal{L}\left[ \hat{a}\right]  \notag \\
&&+\sum\limits_{j=0,1,2}\gamma _{jb}\mathcal{L}\left[ \sigma _{jb}^{-}\right]
+\sum\limits_{j=0,1}\gamma _{j2}\mathcal{L}\left[ \sigma _{j2}^{-}\right]
+\gamma _{01}\mathcal{L}\left[ \sigma _{01}^{-}\right]  \notag \\
&&+\sum\limits_{j=0,2,b}\gamma _{j,\varphi }\left( \sigma _{jj}\rho \sigma
_{jj}-\sigma _{jj}\rho /2-\rho \sigma _{jj}/2\right) ,
\end{eqnarray}%
where $\sigma _{jb}^{-}=\left\vert j\right\rangle \left\langle b\right\vert
, $\ $\sigma _{j2}^{-}=\left\vert j\right\rangle \left\langle 2\right\vert ,$%
\ $\sigma _{01}^{-}=\left\vert 1\right\rangle \left\langle 0\right\vert ,$\ $%
\sigma _{jj}=\left\vert j\right\rangle \left\langle j\right\vert $, $%
\mathcal{L}\left[ \Lambda \right] =\Lambda \rho \Lambda ^{+}-\Lambda
^{+}\Lambda \rho /2-\rho \Lambda ^{+}\Lambda /2$\ (with $\Lambda =\hat{a}%
,\sigma _{jb}^{-},\sigma _{j2}^{-},\sigma _{01}^{-})$,\ $\kappa $\ is the
decay rate of the cavity,\ $\gamma _{jb}$ is the energy relaxation rate of
the level $\left\vert b\right\rangle $\ for the decay path $\left\vert
b\right\rangle \rightarrow \left\vert j\right\rangle $\ of the ququart ($%
j=0,1,2$), $\gamma _{j2}$\ ($\gamma _{01}$) is the relaxation rate of the
level $\left\vert 2\right\rangle $\ for the decay path $\left\vert
2\right\rangle \rightarrow \left\vert j\right\rangle $\ ($\left\vert
0\right\rangle \rightarrow \left\vert 1\right\rangle $) of the ququart ($%
j=0,1$); $\gamma _{j,\varphi }$\ is the dephasing rate of the level $%
\left\vert j\right\rangle $\ of the ququart ($j=0,2,b$).

The operational fidelity is given by $F=\sqrt{\langle \psi _{\mathrm{id}%
}|\rho |\psi _{\mathrm{id}}\rangle }.$\ Here, $|\psi _{\mathrm{id}}\rangle $%
\ is the ideal output state of Eq. (13), which is achieved under the
effective Hamiltonian (5) without taking into account the system dissipation
and the unwanted couplings; while $\rho $\ is the density operator of the
system for the operation being performed in a realistic situation.\

The typical transition frequency between neighboring levels of a SC flux
ququart can be made as 1 to 20 GHz [83-85]. As an example, we consider $%
\omega _{1b}/2\pi =14.5$ GHz, $\omega _{2b}/2\pi =12.5$ GHz, $\omega
_{0b}/2\pi =13.5$ GHz, $\omega _{12}/2\pi =2.0$ GHz, $\omega _{01}/2\pi =1.0$
GHz, and $\omega _{c}/2\pi =10.5$ GHz. As a result, we have $\Delta
_{2b}/2\pi =2.0$ GHz, $\Delta _{1b}/2\pi =4.0$ GHz, $\Delta _{0b}/2\pi =3.0$
GHz, $\Delta _{01}/2\pi =-9.5$ GHz, $\Delta _{02}/2\pi =-9.5$ GHz, and $%
\Delta _{12}/2\pi =-8.5$ GHz. In addition, with appropriate design of the
flux ququart [86], one can have $\phi _{2b}\sim \phi _{1b}\sim $ $\phi
_{12}\sim 10\phi _{0b}\sim 10\phi _{02}\sim 10\phi _{01},$ where $\phi _{ij}$
represents the dipole coupling matrix element between the two levels $%
\left\vert i\right\rangle $\ and $\left\vert j\right\rangle ,$\ with $ij\in
\left\{ 2b,1b,12,0b,02,01\right\} $. Since the coupling constant $g_{ij}$
between the cavity and the $\left\vert i\right\rangle $\ $\leftrightarrow $ $%
\left\vert j\right\rangle $ transition of the ququart is proportional to the
$\phi _{ij}$ associated with the two levels $\left\vert i\right\rangle $\
and $\left\vert j\right\rangle $, one has $g_{2b}\sim g_{1b}\sim $ $%
g_{12}\sim 10g_{0b}\sim 10g_{02}\sim 10g_{01}$.\ In the numerical
simulations, we set $g_{2b}=g_{1b}=$ $g_{12}=2\pi \times 120$ MHz and $%
g_{0b}=g_{02}=g_{01}=2\pi \times 12$ MHz, which are readily achievable in
experiments [87].

Other parameters used in the numerical simulations are: (i) $\gamma
_{0b}^{-1}=\gamma _{02}^{-1}=\gamma _{01}^{-1}=5T$ $\mu $s, $\gamma
_{1b}^{-1}=\gamma _{2b}^{-1}=T/2$ $\mu $s, $\gamma _{12}^{-1}=T$\ $\mu $s,
(ii) $\gamma _{b,\phi }^{-1}=\gamma _{2,\phi }^{-1}=T/2$\ $\mu $s, $\gamma
_{0,\phi }^{-1}=2.5T$, and (iii) $\alpha =3.05.$\ For $T=$\ $20$\ $\mu $s,
the decoherence times of the SC flux ququart used in the numerical
simulations are $10$\ $\mu $s$-100$ $\mu $s, which is a rather conservative
case as the decoherence time 70 $\mu $s to 1 ms for a superconducting flux
device has been experimentally demonstrated [52,88]. Furthermore, a cat
state with $\alpha =3.05$ can be created in experiments because the
circuit-QED experiments have generated a cat state with an amplitude $\alpha
\leq 5.27$ [78-82].

\begin{figure}[tbp]
\begin{center}
\includegraphics[width=9.5 cm, clip]{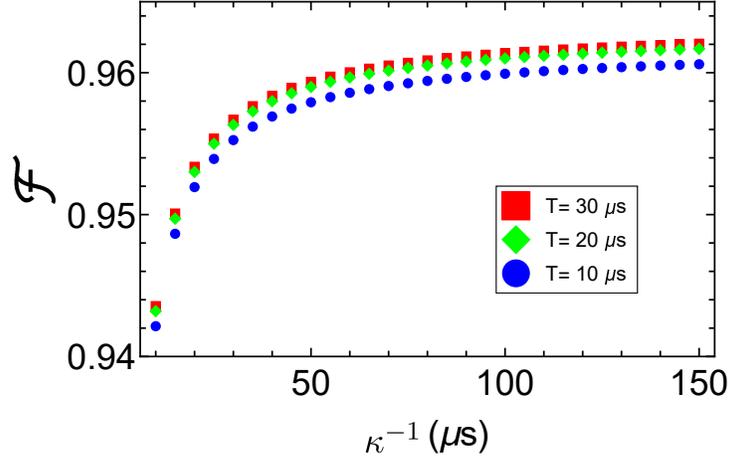} \vspace*{-0.08in}
\end{center}
\caption{(color online) Fidelity versus $\protect\kappa ^{-1}$ for $T=10~%
\protect\mu $s, $20~\protect\mu $s, and $30~\protect\mu $s.}
\label{fig:3}
\end{figure}

\begin{figure}[tbp]
\begin{center}
\includegraphics[width=9.5 cm, clip]{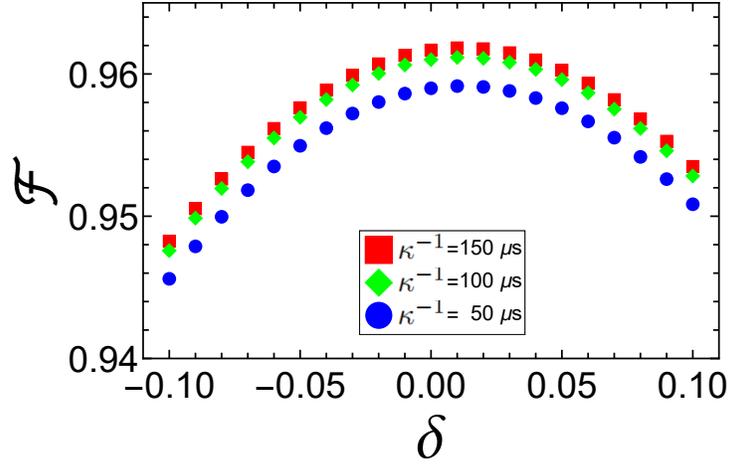} \vspace*{-0.08in}
\end{center}
\caption{(color online) Fidelity versus $\protect\delta $ for $\protect%
\kappa ^{-1}=50~\protect\mu $s, $100~\protect\mu $s, and $150~\protect\mu $%
s. We set $T=$ $20\protect\mu $s in the numerical simulation.}
\label{fig:4}
\end{figure}

By numerically solving the master equation (15), Fig. 3 is plotted to show
the fidelity versus $\kappa ^{-1}$\ for $T=10~\mu $s, $20~\mu $s, and $%
30~\mu $s. Figure 3 shows that the fidelity exceeds $96.1\%$ for $\kappa
^{-1}\geq 100$\ $\mu $s and $T\geq 20~\mu $s. Furthermore, Fig. 3
demonstrates that the fidelity is sensitive to the cavity decay and the
decoherence of the ququart. This can be easily understood since the photons
are populated in the cavity and both levels $\left\vert 0\right\rangle $ and
$\left\vert 2\right\rangle $ of the ququart are occupied during the
preparation of the hybrid entangled state (13).

In practice, the initial state of Eq.~(12) may not be perfectly prepared.
Thus, we consider a non-ideal initial state of the system%
\begin{eqnarray}
\left\vert \psi \left( 0\right) \right\rangle _{\mathrm{non-ideal}} &=&%
\mathcal{N}_{1}^{-1}\left[ \left( 1/\sqrt{3}+\delta \right) |0\rangle +1/%
\sqrt{3}|1\rangle +\left( 1/\sqrt{3}-\delta \right) |2\rangle \right]  \notag
\\
&&\times \mathcal{N}_{2}\left( \sqrt{1+\delta }|\alpha \rangle +\sqrt{%
1-\delta }|-\alpha \rangle \right) ,
\end{eqnarray}%
where $\mathcal{N}_{1}=\sqrt{1+2\delta ^{2}}$ and $\mathcal{N}_{2}=1/\sqrt{%
2+2\sqrt{1-\delta ^{2}}e^{-2\left\vert \alpha \right\vert ^{2}}}.$ In this
case, we numerically plot Fig. 4 for $T=20~\mu $s. Figure 4 shows that the
fidelity decreases with increasing $\delta $. In addition, Fig. 4 shows that
for $\delta \in \lbrack -0.1,0.1],$\ namely, a $10\%$\ error in the weights
of $|\alpha \rangle $\ and $|-\alpha \rangle $\ states as well as in the
weights of $|0\rangle $\ and $|2\rangle $, a fidelity greater than $94.5\%,$
$94.7\%,$ and $94.8\%$ can be obtained for $\kappa ^{-1}=50~\mu $s, $100~\mu
$s, $150~\mu $s, respectively.

The operational time for preparing the hybrid entangled state (13) is
estimated to be $\sim 0.046$\ $\mu $s, much shorter than the cavity decay
time 10 $\mu $s$-$150 $\mu $s applied in the numerical simulations. For the
cavity frequency given above and $\kappa ^{-1}=100~\mu $s, the quality
factor of the cavity is $Q\sim 6.59\times 10^{6}$, which is available
because a 3D microwave cavity with a high quality factor $Q=3.5\times 10^{7}$
was experimentally reported [89,90].

\begin{center}
\textbf{C. Discussion}
\end{center}

The above analysis shows that the operational fidelity is sensitive to
errors in the initial state preparation, the cavity decay, and the
decoherence of the SC ququart. Numerical simulations indicate that a high
fidelity can still be obtained as long as the error in the initial state
preparation is small. To obtain a high fidelity, it is necessary to reduce
the error in the initial state prepare, select the cavity with a high
quality factor, and use the SC ququart with a long coherence time. The
fidelity can also be improved by employing the SC flux ququart with greater
energy level anharmonicity, such that the unwanted couplings of the cavity
with irrelevant level transitions of the SC flux ququart are negligible.
Lastly, it should be remarked that further studies are needed for each
particular experimental setup. However, this requires a rather lengthy and
complex analysis, which is beyond the scope of this theoretical work.

\begin{center}
\textbf{VI. CONCLUSIONS}
\end{center}

We have proposed an approach to implement a hybrid two-qutrit CSUM gate with
one SC qutrit controlling a target cat-state qutrit. The gate is realized by
the dispersive coupling of the cavity to the SC ququart. As shown above,
this proposal has the following features and advantages: (i) The gate
realization is quite simple, because it requires only a single basic
operation; (ii) Neither classical pulse nor measurement is needed; (iii) The
hardware resources are minimized because no auxiliary system is required for
the gate implementation; and (iv) The auxiliary higher energy level of the
SC ququart is virtually excited during the gate operation, thus decoherence
from this level is greatly suppressed. To our knowledge, this work is the
first to demonstrate the realization of the proposed hybrid gate based on
cavity or circuit QED. This proposal is quite general and can be applied to
implement a hybrid two-qutrit CSUM gate with one matter qutrit (of different
type) controlling a target cat-state qutrit in a wide range of physical
system, such as a four-level natural atom or artificial atom (e.g., a
quantum dot, an NV center, a SC ququart, etc.) coupled to an optical or
microwave cavity.

As an application, we have further discussed the generation of a hybrid
maximally-entangled state of a SC qutrit and a cat-state qutrit, by applying
the proposed hybrid gate. To our knowledge, this work is the first to show
the preparation of a hybrid entangled state of a SC qutrit and a cat-state
qutrit. We have also numerically analyzed the experimental feasibility of
generating such hybrid entangled state within current circuit QED
technology. We hope that this work will stimulate experimental activities in
the near future.

\begin{center}
\textbf{ACKNOWLEDGEMENTS}
\end{center}

This work was partly supported by the National Natural Science Foundation of
China (NSFC) (11074062, 11374083, 11774076, U21A20436), the Jiangxi Natural
Science Foundation (20192ACBL20051), and the Key-Area Research and
Development Program of GuangDong province (2018B030326001).

\end{document}